\documentclass[english,amsmath,amssymb,pra,aps,showpacs,mathbbm,superscriptaddress,twocolumn]{revtex4-1}
\usepackage{amsmath,amsfonts,amssymb,amsthm,graphics,graphicx,epsfig,times,bbm}
\usepackage[colorlinks=true,citecolor=blue,linkcolor=blue,urlcolor=blue]{hyperref}
\usepackage[usenames]{color}
\usepackage{graphicx}
\usepackage{subfigure}
\usepackage{amsmath}
\usepackage{epsfig,times}
\usepackage{dcolumn}
\usepackage{bm}
\usepackage{color}
\usepackage{verbatim}
\usepackage{epstopdf}
\usepackage{amssymb}
\usepackage{amstext}
\usepackage{latexsym}
\usepackage{hyperref}
\usepackage{amsfonts}
\usepackage{psfrag}
\usepackage{xcolor}

\makeatletter

\usepackage{amsfonts}
\usepackage{epsfig}
\usepackage{dsfont}

\newcommand{\be}{\begin{equation}}
\newcommand{\bea}{\begin{eqnarray}}
\newcommand{\eea}{\end{eqnarray}}
\newcommand{\ee}{\end{equation}}

\newcommand{\quot}[1]{``#1''}

\newcommand{\C}{{\cal C}}

\newcommand{\N}{{\cal N}}
\newcommand{\A}{{\cal A}}
\newcommand{\G}{{\cal G}}
\newcommand{\bA}{{\boldsymbol{A}}}

\def\H{{\cal H}}

\usepackage{babel}
\makeatother

\begin{document}

\title{Uncontrolled disorder effects in fabricating photonic quantum simulators on a kagome geometry: A 
projected-entangled pair state versus exact digonalization
analysis}

\author{Amin \surname {Hosseinkhani}}
\affiliation{Department of Physics, Shahid Beheshti University, G.C., Evin, Tehran 19839, Iran}

\author{Bahareh \surname {Ghannad Dezfouli}}
\affiliation{Department of Optics, Faculty of Science, Palack\'{y} University, 17. listopadu 12, 77146 Olomouc, Czech Republic}
\affiliation{Department of Physics, Shahid Beheshti University, G.C., Evin, Tehran 19839, Iran}

\author{Fatemeh \surname {Ghasemipour}}
\affiliation{Department of Physics, Shahid Beheshti University, G.C., Evin, Tehran 19839, Iran}

\author{Ali T. \surname {Rezakhani}}
\affiliation{Department of Physics, Sharif University of Technology, Tehran 14588, Iran}

\author{Hamed \surname {Saberi}}
\email{saberi@optics.upol.cz}
\affiliation{Department of Optics, Faculty of Science, Palack\'{y} University, 17. listopadu 12, 77146 Olomouc, Czech Republic}
\affiliation{School of Physics, Institute for Research in Fundamental Sciences (IPM), Tehran 19395-5531, Iran}
\affiliation{Department of Physics, Shahid Beheshti University, G.C., Evin, Tehran 19839, Iran}

\date{June 27, 2014}

\begin{abstract}

 We propose a flexible numerical framework for extracting the energy spectra and photon transfer dynamics of a unit kagome cell with disordered cavity-cavity couplings under realistic experimental conditions. A projected-entangled pair state (PEPS) ansatz to the many-photon wavefunction allows to gain a detailed understanding of the effects of undesirable disorder in fabricating well-controlled and scalable photonic quantum simulators. The correlation functions associated with the propagation of two-photon excitations
 reveal intriguing interference patterns peculiar to the kagome geometry and promise at the same time a highly tunable quantum interferometry device with a signature for the formation of resonant or Fabry-Pe\'rot-like transmission of photons. Our results justify the use of the proposed PEPS technique for addressing the role of disorder in such quantum simulators in the microwave regime and promises a sophisticated numerical machinery for yet further explorations of the scalability of the resulting kagome arrays. The introduced methodology and the physical results may also pave the way for unraveling exotic phases of correlated light on a kagome geometry.


\end{abstract}



\maketitle

\section{Introduction and the model}
\label{sec:intro}

\begin{figure}[t]
\centering
\includegraphics[width=1.0\linewidth]{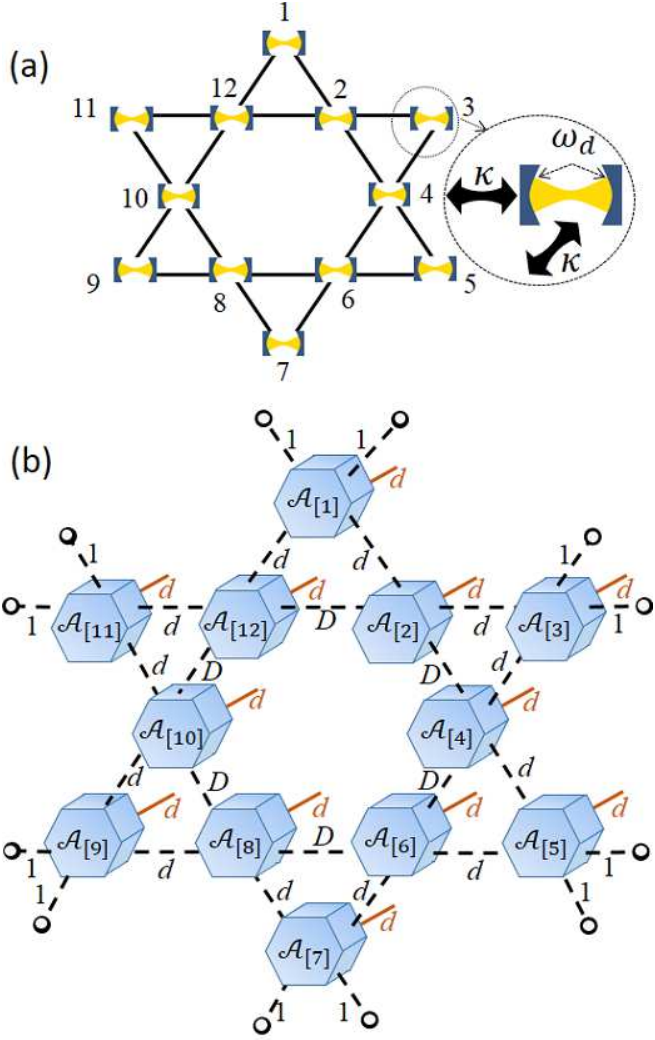}
\caption{(Color online) (a) Real-space kagome photon cell takes the OBC tensor network representation in (b). The hexagonal prisms represent the PEPS $\A$-tensors in Eq.~(\ref{eq:PEPS_ansatz}) each possessing one \emph{physical} index $i_k$ (shown by solid line) of dimension $d$ and four \emph{virtual} indices (dashed lines) of dimension 1, $d$, or $D$ depending where it sits in the kagome. The local Hilbert space dimension for a total number of $N$ photons is $d = N+1$, as many as all possible bosonic excitations each cavity can accommodate, i.e., $n_k \in \{0, 1, \cdots, N\}$ where $n_k$ is the number of photons in cavity $k$. The exact (or truncated) $N$-photon Hilbert space dimension is denoted by $D$.}
\label{fig:kagome_PEPS}
\end{figure}

The idea of employing well-controlled quantum systems to \quot{simulate} complex quantum matter was first put forward by Richard Feynman in a keynote speech in 1981~\cite{Feynman1982}. Since then a lot of efforts has been devoted to propose various physical setups and scenarios for putting into practice such \emph{quantum simulators} that promise to address otherwise intractable problems of Nature. Of particular interest is the potential application of such \quot{problem-solvers} operating at the quantum level to efficiently reproduce the dynamics of other many-body quantum systems. Among various candidates for physical implementation of quantum simulation, the idea of using \emph{photons} as particles in a quantum simulator has received growing attention in recent years due to the flexibility afforded by lithographic fabrication and the relative ease of achieving strong coupling within a
superconducting circuit architecture~\cite{Houck2012}. The \emph{cavity lattices} so fabricated, in particular, provide a versatile testbed and viable platform for quantum simulation of strongly correlated systems both in and out of equilibrium~\cite{Houck2012}. They have been conjectured to harbor a wide spectrum of collective phenomena, such as a superfluid{\textendash}Mott-insulator transition~\cite{Hartmann2006,Greentree2006,Koch2009}, fermionization of photons~\cite{DSouza2013},
anomalous quantum Hall effects~\cite{Petrescu2012}, just to name a few.
Such lattices comprise arrays of coupled on-chip microwave resonators in a \emph{kagome} geometry [see Fig.~\ref{fig:kagome_PEPS}(a)] as the most natural two-dimensional geometry for such transmission line resonators~\cite{Schmidt2010,Houck2012,Underwood2012}.
We remind that the kagome geometry \emph{per se}, is home to a variety of exotic physical phenomena and has sparked an active line of research to address unconventional phases of light and matter on such a geometry~\cite{Yan2011,Petrescu2012}.

A quantitative analysis of microwave cavity lattices has hitherto been possible only for a small number of cavities~\cite{Angelakis2007}, in one spatial dimension~\cite{Hartmann2012}, and with brute force diagonalization techniques~\cite{Underwood2012}. However, more sophisticated and efficient numerical techniques that can account for many-polariton correlations are needed in order to study the dynamics of larger arrays with the possibility of the emergence of various intriguing collective phenomena and phase transitions of light~\cite{Koch2009,Schiro2012,Zheng2011}. In this work, we propose, instead, a combination of the exact diagonalization (ED) technique and the natural generalization of matrix-product states (MPS)~\cite{Schoen2005,Perez2007,Saberi2009,Saberi2012} to two spatial dimensions, i.e., the projected entangled-pair states (PEPS)~\cite{Verstraete2008,Schuch2010} as a flexible numerical machinery to capture various static and dynamic properties of an exemplifying photonic simulator setup. The results promise numerical capabilities for exploring yet larger arrays.

Moreover, engineering identical couplings between cavities is a task of formidable
difficulty in an experiment; hence, some level of randomness in the couplings needs to be considered within a realistic scenario~\cite{Underwood2012}. Such an uncontrolled disorder poses major obstacle to the functionality of the kagome arrays as well-controlled quantum simulators. A quantitative assessment of the effect of disorder is thus essential for analyzing the feasibility of quantum simulation in such lattices. The presence of randomness in the model, on the other hand, hampers an analytical investigation via common continuous Fourier analysis as we shall elaborate in the subsequent section. Our sophisticated PEPS approach, instead, is capable of addressing the model even under the assumption of disordered couplings.

As an important step toward the realization of photonic quantum simulators, we consider here a cavity lattice consisting of a unit kagome cell described by a bosonic tight-binding Hamiltonian of the form
\begin{eqnarray}
\label{eq:tight-bind}
\hat{\H} = \hbar \omega_d \sum_{k=1}^{12} \hat{a}_k^{\dagger} \hat{a}_{k} - \kappa \sum_{\langle k,k' \rangle} (\hat{a}_k^{\dagger} \hat{a}_{k'} + \mathrm{H.c.}) \; ,
\end{eqnarray}
where $\omega_d$ is the driving frequency, $\hat{a}_k^{\dagger} (\hat{a}_k)$ is the photon creation (annihilation) operator for resonator $k$, and $\kappa$ denotes the hopping strength between the nearest-neighbor cavities. According to (\ref{eq:tight-bind}), photons from a microwave source are injected into one of the twelve cavities of the kagome cell and are able to hop into other cavities with a hopping strength that can be tuned in an experiment~\cite{Underwood2012}. The competition between the field energy and the hopping decides the phase diagram of the model.

It is noteworthy to mention that the hopping Hamiltonian~(\ref{eq:tight-bind}) may alternatively be interpreted as a Jaynes-Cummings-Hubbard (JCH) type of Hamiltonian~\cite{Angelakis2007} in the absence of a two-level system inside the cavity. Nonetheless, it should be emphasized that although the JCH model provides a paradigm of realizing an effective interaction between photons, a resonant two-level system (the local Jaynes-Cummings interaction inside each cavity) is not a requirement for realizing effective photon-photon interaction, since strong off-resonant interactions have also been observed in coupled cavity lattices~\cite{Hoffman2011}, not to mention that fabrication of a JCH based photonic simulator lies far from the current experimental reach~\cite{Underwood2012}.


Furthermore, although the Hamiltonian model (\ref{eq:tight-bind}) might seem to be analytically diagonlizable in the momentum space, but the presence of disorder in cavity-cavity couplings breaks the required translational invariance for a continuous Fourier analysis and even a straightforward extraction of physical quantities (e.g., ground state energy) via a discrete Fourier transform version faces daunting challenges in two spatial dimension. Our proposed PEPS approach, instead, is not subject to such technical restrictions.

A PEPS ansatz to the many-photon Hamiltonian of Eq.~(\ref{eq:tight-bind}) brings about a highly flexible structure to easily access the eigenspectrum and efficiently calculate other expectation values and various correlation functions. Such an ansatz to the many-photon ground state of a single \emph{closed} kagome with open-boundary condition (OBC) is given by
\begin{eqnarray}
\label{eq:PEPS_ansatz}
\hspace{-4mm} |\Psi_{\mathrm{G}}\rangle = \sum_{i_1, i_2 ,\cdots, i_{12} = 1}^d
\C(\A_{[1]}^{i_1}, \A_{[2]}^{i_2}, \cdots, \A_{[12]}^{i_{12}})
\bigotimes_{k=1}^{12}  |i_k\rangle \; ,
\end{eqnarray}
where the PEPS coefficients $\C (\A_{[1]}^{i_1}, \cdots, \A_{[12]}^{i_{12}} )$ are the outcome of the \quot{contractions} of the virtual indices pertaining to the PEPS $\A$-tensors of rank 5, and $|i_k\rangle$ denotes the local Hilbert space of dimension $d$ at cavity $k$. We stress a finite closed system without periodic boundary condition is considered in this work. The contraction scheme denoted by the function $\C(.)$ follows the underlying kagome lattice structure with an associated tensor network representation depicted in Fig.~\ref{fig:kagome_PEPS} (b). We shall henceforth refer to (\ref{eq:PEPS_ansatz}) as a \emph{kagome PEPS}. Tensor network ans\"atze~\cite{Orus2013} provide powerful tools for the study of quantum many-body systems in the low-energy regime by representing the state of a system as an efficiently-contractible network of multi-index tensors optimized numerically by means of an \textit{in situ} variational algorithm~\cite{Verstraete2004,Saberi2008}. We propose in the following a tensor network algorithm for capturing the ground state of the kagome cell described by Hamiltonian~(\ref{eq:tight-bind}).

\begin{figure}[t]
\centering
\includegraphics[width=1.0\linewidth]{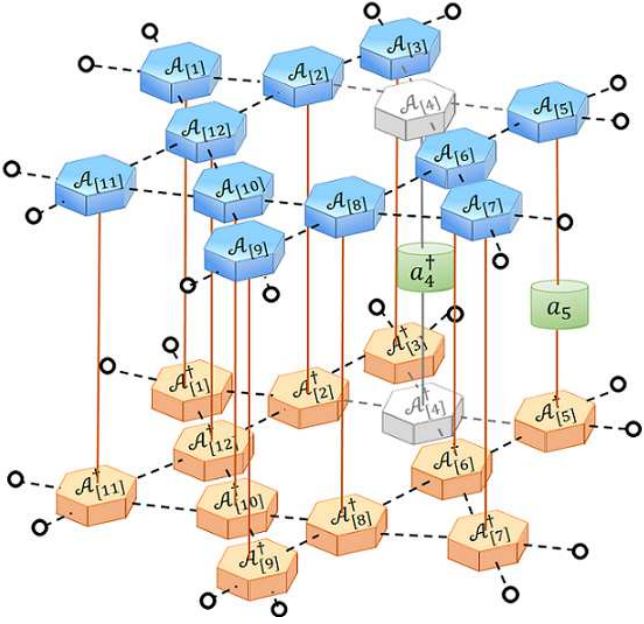}
\caption{(Color online) Tensor network representation of $\H_k^{\mathrm{eff.}}$ in Eq.~(\ref{eq:effective_H_C}a) upon calculating the expectation value of an exemplifying hopping term $\hat{a}^{\dagger}_4 \hat{a}_5$ at $k=4$. The contraction is carried out over all connected indices excluding those associated with the \quot{current} tensor $\A_{[4]}$.}
\label{fig:hopping}
\end{figure}

\section{The PEPS variational optimization for kagome}
\label{sec:VPEPS_kagome}

The ground state $|\Psi_{\mathrm{G}}\rangle$ of the many-photon Hamiltonian (\ref{eq:tight-bind}) is recognized as the trial wavefunction $|\Psi\rangle$ that minimizes the expectation value
\begin{eqnarray}
\label{eq:variation}
\min_{|\Psi \rangle \in \{\mathrm{PEPS}\} } \dfrac{\langle \Psi | \hat{\H} | \Psi \rangle}{\langle \Psi | \Psi \rangle} \; .
\end{eqnarray}
The minimization can be carried out efficiently using a \quot{sweeping procedure} in which one fixes all but the $k$'th PEPS $\A$-tensor and iteratively optimizes it while contracting all the indices in the numerator and denominator of Eq.~(\ref{eq:variation}) save those connecting to $\A_{[k]}$ and $\A^{\dagger}_{[k]}$. By interpreting the tensor $\A_{[k]}$ as a $(d \times \prod_{i=1}^4 D_i^{\mathrm{v}})$-dimensional vector ${\bA}_k$ ($D_i^{\mathrm{v}}$ denoting the virtual index dimension), these expressions can be written as~\cite{Verstraete2004,Verstraete2008}
\begin{subequations}
\label{eq:effective_H_C}
\begin{eqnarray}
\langle \Psi | \hat{\H} | \Psi \rangle & = & {\bA}_k^{\dagger} \H_k^{\mathrm{eff.}} {\bA}_k , \\
\langle \Psi | \Psi \rangle & = & {\bA}_k^{\dagger} \N_k^{\mathrm{eff.}} {\bA}_k \; ,
\end{eqnarray}
\end{subequations}
where $\H_k^{\mathrm{eff.}}$ and $\N_k^{\mathrm{eff.}}$ are called the \quot{effective Hamiltonian} and the \quot{effective normalization} at site $k$, respectively [see Fig.~\ref{fig:hopping}]. Thus, the minimization (\ref{eq:variation}) translates into a \textit{generalized eigenvalue problem} of the form
\begin{eqnarray}
\label{eq:gen_eig}
\H_k^{\mathrm{eff.}} {\bA}_k =  \xi_k \N_k^{\mathrm{eff.}} {\bA}_k \; ,
\end{eqnarray}
with the smallest generalized eigenvalue so obtained to be recognized as the optimized upper bound to the ground state energy $E_{\mathrm{G}} \le \min_j{\xi_k^{j}} \equiv \xi_k^{\mathrm{min}}$. After reshaping back the corresponding generalized eigenvector $\bA_k^{\mathrm{min}}$ to the new optimal tensor, say ${\tilde{\A}}_{[k]}$, we proceed to the next site, and iterate such a procedure by sweeping through the whole kagome until convergence of $\xi_k^{\mathrm{min}}$ (the best estimate for the ground state energy) is achieved.

The generalized eigenvalue problem of Eq.~(\ref{eq:gen_eig}) remains well-conditioned as long as $\N_k^{\mathrm{eff.}}$ stays nonsingular. In one-dimensional variational MPS with OBC one can always orthonormalize the tensor network in such a way that $\N_k^{\mathrm{eff.}}$ boils down to an identity matrix and the numerical stability is thereby guaranteed by construction~\cite{Saberi2008}. In two-dimensional problems, on the contrary, the spectrum of $\N^{\mathrm{eff.}}_k$ might generically contain ill-disposed $\xi_k$'s below the numerical precision. In light of Eqs.~(\ref{eq:variation})--(\ref{eq:gen_eig}), we propose then the deviation
\begin{eqnarray}
\label{eq:error}
 \xi_k^{\mathrm{min}}- \frac{{\bA_k^{\mathrm{min}}}^{\dagger} \H_k^{\mathrm{eff.}} \bA_k^{\mathrm{min}}} {{\bA_k^{\mathrm{min}}}^{\dagger} \N_k^{\mathrm{eff.}} \bA_k^{\mathrm{min}}} \; ,
\end{eqnarray}
as the figure of merit that can signal such an ill-conditioning throughout the whole simulation.

\begin{figure}[t]
\centering
\includegraphics[width=1.0\linewidth]{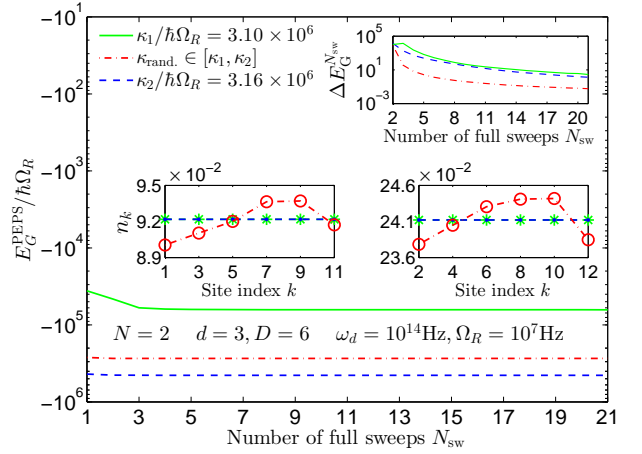}
\caption{(Color online) The convergence of the ground state energy out of PEPS $E_{\mathrm{G}}^{\mathrm{PEPS}}$  as well as the effect of disorder in the hopping strengths $\kappa$. The upper inset shows the \quot{convergence speed} by plotting the difference at subsequent sweeps $\Delta E_{\mathrm{G}}^{N_{\mathrm{sw}}} \equiv E^{\mathrm{PEPS}}_{\mathrm{G}}(N_{\mathrm{sw}})- E^{\mathrm{PEPS}}_{\mathrm{G}}(N_{\mathrm{sw}}-1)$ versus the full sweep number $N_{\mathrm{sw}}$. Equilibrium local occupation numbers are also provided in the lower insets. Each \quot{full sweep} consists of a clockwise optimization of the PEPS $\A$-tensors followed by a counterclockwise one.}
\label{fig:sweeping}
\end{figure}

\section{Tuning the total number of photons in the kagome}
\label{sec:chemical_potential}

The total number of photons in the kagome $N$ can be fixed by tuning the ratio $(\hbar\omega_d\ -\mu)/ \kappa$ in a grand-canonical description in which the Hamiltonian $\hat{\H}$ is replaced by $ \hat{\H} - \mu \hat{N}$. In the latter, $\mu$ denotes the \quot{chemical potential}, and $\hat{N} \equiv \sum_{k} \hat{a}_{k}^{\dagger} \hat{a}_k$
is the operator for the total number of excitations. It must be noted that although a pure photonic lattice is considered in the present work and photons in general exhibit a zero chemical potential, here a $\mu$ term is simply used as a numerical tool to control the mean cavity occupancy and for working in a fixed-$N$ sector of the many-photon Hilbert space~\cite{Bardyn2012,Schiro2012}. Although practical recipes for engineering
such an effective chemical potential in a real experiment need to be developed yet~\cite{Houck2012}, a possible proposal might be formulated based on a control procedure via an interplay of photon input and loss rates. In this sense the
chemical potential is purely a theoretical construct here, as opposed to electronic systems where it is a real potential (e.g., an applied voltage).

Despite the success of the recipe for fixing the number of a few photonic excitations, we have found out in practice that the method poses numerical restrictions by yielding increasingly narrower fixed-$N$ intervals of $\kappa$ for larger number of photons. A possible remedy may be realized upon implementing the Abelian $U(1)$ symmetry~\cite{Bauer2011} in the representation of the Hamiltonian~(\ref{eq:tight-bind}) that inherently conserves the total number of excitations. Another possible strategy might parallel the use of Lagrange multipliers for implementing the constant-norm constraint in MPS algorithms~\cite{Saberi2008} and its possible extension for preserving the number of photons in the kagome PEPS.

\section{Results for the equilibrium properties of the kagome photon cell}
\label{sec:ground_state}

Figure~\ref{fig:sweeping} illustrates the results of the application of the outlined PEPS procedure for capturing the ground state of the kagome with $N=2$. The energy has been normalized with respect to $\hbar \Omega_R$, where $\Omega_R$ denotes the Rabi frequency associated with a typical microwave circuit quantum electrodynamics (QED) value of $\Omega_R = 10^7 \mathrm{Hz}$ in the strong coupling regime~\cite{Wallraff2004} and in the interest of prospective applications to an \emph{array} of circuit QED on a kagome geometry. Nonetheless, a quantitative scaling of the results to arbitrary values of driving frequencies (e.g., optical range) may easily be understood by multiplying the related values by a ratio of $\Omega_R/\omega_d$.

We stress, although $\Omega_R$ does not appear in the Hamiltonian explicitly, it already sets an energy scale that shall profit further circuit-QED developments in strong or ultrastrong coupling regime defined by the ratio of the qubit-cavity coupling to such a frequency~\cite{Niemczyk2010,Romero2012,Felicetti2013}.

The main plot illustrates the convergence of the ground state energy of such a model under the sweeping procedure and starting from a randomly initialized kagome PEPS. The convergence is satisfactorily reached shortly after a few sweeps. The role of disorder in cavity-cavity couplings has been analyzed by a random selection of $\kappa_{\mathrm{rand}}$'s from an illustrative interval of fixed-$N$ $[\kappa_1, \kappa_2]$ giving rise to the values of energies that are bounded from above (below) by those of $\kappa_1$ ($\kappa_2$).


The lower insets of Fig.~\ref{fig:sweeping}, furthermore, show the equilibrium population of photons at site $k$ obtainable from
\begin{eqnarray}
\label{eq:local_occupation}
n_k = \langle \Psi_{\mathrm{G}} | \hat{a}_k^{\dagger} \hat{a}_{k} | \Psi_{\mathrm{G}} \rangle \; ,
\end{eqnarray}
with the converged value of $|\Psi_{\mathrm{G}} \rangle$ to be used in the latter while exploiting similar graphical recipes as that of Fig.~\ref{fig:hopping} for calculating the expectation values. From the results so obtained the equilibrium and energetically favorable distribution of photons in the kagome can be inferred: The photons get uniformly distributed on either the inner (even $k$) or outer (odd $k$) parts of the kagome with a higher probability associated with the inner sites and in compliance with the constraint $N = 6 (n_{\mathrm{inner}} + n_{\mathrm{outer}})$ that is rooted in the kagome symmetry. The relatively lower values of $n_{\mathrm{outer}}$, on the other hand, are a consequence of \quot{photon reflection} at the boundaries. The presence of disorder in the couplings, however, breaks the latter inner/outer symmetry, as the open circles depict in the same insets.

\begin{figure}[t]
\centering
\includegraphics[width=1.0\linewidth]{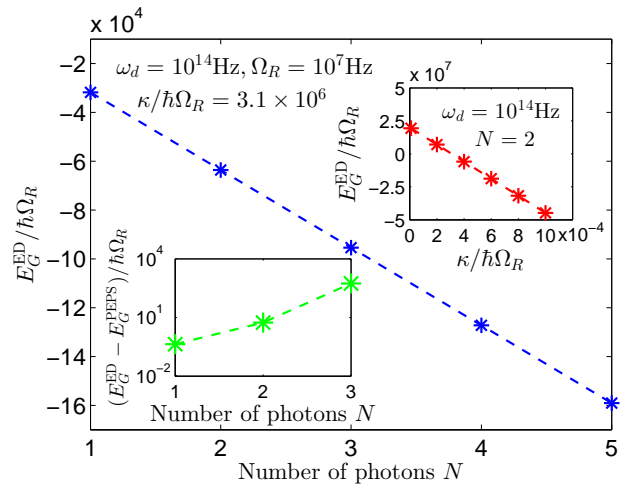}
\caption{(Color online) Scaling of the ground state energy of the kagome cell with the hopping strength and the number of photons. The upper inset shows the data for the ground state energy of two photons with respect to the hopping strength $\kappa$ obtained from the ED method. The main plot illustrates the scaling of the ground state
energy with an illustrative number of up to $N = 5$ photons obtained
from ED. In the lower inset, the difference of the ground state energies out of PEPS and ED
are plotted. In the latter, a cutoff of $D=6$ on the PEPS bond dimension has been used for $N=3$ data.}
\label{fig:N_kappa_scaling}
\end{figure}

Figure~\ref{fig:N_kappa_scaling} explores the scaling of the ground state energy with the total number of photons $N$ and the hopping strength $\kappa$. A linear scaling of the ground-state energy with the hopping strength $κ$ is illustrated in the upper inset of the figure for an illustrative value of $N=2$ and shall be exploited later on for interpreting the results for the dynamic properties of the kagome in the subsequent section.

The main plot suggests a linear scaling of the ground state energy with an illustrative number of up to $N = 5$ photons obtained from the exact diagonalization of the model within each fixed-$N$ subspace. The relatively higher value
of $N$ compared to the one in Fig.~\ref{fig:sweeping} has to do with less memory demands on the side of ED.

The lower inset gives the difference of the results for the ground state energies obtained from PEPS and ED. A good agreement
up to $N =2$ photons confirms the success of the proposed PEPS approach in capturing the ground state of few-photon kagome cell. It must however be noted that for $N=2$ a better agreement might yet be achieved if one allows for yet further optimization sweeps requiring, in turn, much longer CPU time due to its taking advantage of a bond dimension as high as $D=9$. The rather substantial disagreement between the PEPS and ED for $N=3$ stems from the use of a PEPS truncation on the size of the PEPS tensors down to those with $D=6$. One may note that the exact PEPS bond dimension in this case and following the graphical pattern in Fig.~\ref{fig:kagome_PEPS}(b) would have been as high as $D=d^2=(N+1)^2=16$ leading to PEPS $\A$-tensors of intractable sizes within our available memory resources. Applying a geometrical cutoff on the PEPS bond dimension hence becomes inevitable for making things tractable in that direction. Yet, the PEPS variational optimization is supposed to realize the best possible (approximate) description of the ground state within the available and truncated resources. 

We point out moreover our aim in the lower inset has been to benchmark our proposed PEPS approach by using the largest possible value of $N$ on the PEPS side and within our numerical resources, i.e., $N=3$, in terms of the memory demands associated with storing huge PEPS $\A$-tensors.


Some remarks on a balanced appraisal of the relative pros and cons of the PEPS and ED are in order:

(i) The ED method is based on a brute force construction of the fixed-$N$ initial \emph{subspace} by exhausting all possibilities of the distribution of $N$ photons in the kagome which is feasible for a small number of photons but becomes increasingly intractable for larger ones. In contrast, the PEPS approach starts from a randomly initialized many-photon state and variationally searches for the eigenspectrum of the model within the \emph{full} Hilbert space.

(ii) PEPS can help address large \emph{arrays} of kagome since it allows harnessing the size of the many-body Hilbert space by introducing a cutoff on the PEPS bond dimension $D$ whereas ED lacks such a structural capability.

(iii) Unlike the rigid and inflexible method of ED, PEPS is variational in nature and allows \emph{local} access to the information stored in the PEPS $\A$-tensors and their further optimization at will. This could, for example, provide vivid insights into the \emph{entanglement spectrum} of the model through a local construction of the reduced density matrix at each site and a straightforward calculation of the \emph{Schmidt gap} as the recently introduced indicator of a quantum phase transition in many-body systems~\cite{DeChiara2012}.

\section{Results for the dynamic properties of the kagome photon cell}
\label{sec:dynamic}

\begin{figure}[t]
\centering
\includegraphics[width=1.0\linewidth]{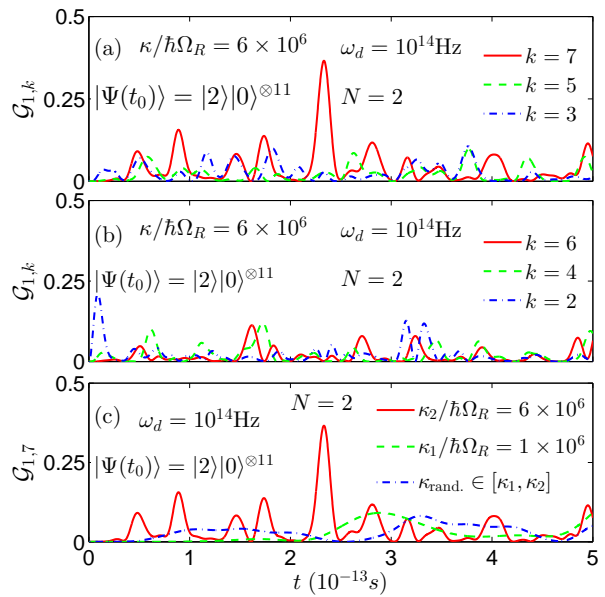}
\caption{(Color online) Two-point correlation functions of Eq.~(\ref{eq:corr_func}) obtained via the ED of the kagome photon cell. The kagome starts out in a Fock state with two photons localized in the arbitrarily chosen reference cavity $k=1$.}
\label{fig:corr_func}
\end{figure}

We finally address the real-time dynamics of the kagome by calculating the two-point correlation function of the form
\begin{eqnarray}
\label{eq:corr_func}
\G_{k, k'}(t) \equiv \langle \Psi(t) | \hat{n}_k \hat{n}_{k'} | \Psi(t) \rangle \; ,
\end{eqnarray}
where the state of kagome at time $t$ ($|\Psi(t)\rangle$) is obtained via the action of the unitary time evolution operator $\hat{U}(t_0, t) \equiv e^{-i \hat{\H} t/ {\hbar}}$ on some initial state of the kagome $|\Psi(t_0)\rangle$.
The correlation function so defined evaluates the average result of a joint photon-number measurement performed on cavities $k$ and $k'$. The time evolution operator can be calculated from the knowledge of the eigenspectrum of $\hat{\H}$ within the fixed-$N$ subspace of the model.

Figure~\ref{fig:corr_func} shows the two-point correlation functions associated with the propagation of a \emph{localized} two-photon excitation. Assuming $k=1$ as the reference cavity, the \quot{dialogue} with the right half of the kagome is plotted in Figs.~\ref{fig:corr_func}(a) and \ref{fig:corr_func}(b). Note that exactly the same trend keeps repeating in the left half of the kagome due to the kagome's symmetry. The strongest dialogue occurs between the reference cavity and the one at $k'=7$ and may be understood in terms of a \emph{constructive interference} of clockwise and counterclockwise propagating photons with identical optical path differences from the two cavities. The correlation is significantly suppressed for $k' \ne 7$ owing to a \emph{destructive} interference of clockwise and counterclockwise waves that have traversed different optical paths and arrived \emph{out of phase} in either cavities. The peak of $\G_{1,7}$ in (a) associated with the first constructive interference of the photons keeps repeating in regular time intervals due to the kagome symmetry. In Fig.~\ref{fig:corr_func}(c) 
the effect of disorder in hopping strength $\kappa$ on the correlation pattern has been investigated. The randomness in $\kappa$ smears out the sharp communication profile between the cavities. Shorter period of oscillations upon increasing $\kappa$ is also apparent in the plot and may be associated with the increasingly higher energies of photons [c.f., upper inset of Fig.~\ref{fig:N_kappa_scaling}] that, in turn, lead to a faster flow of information.
We point out, the latter shows that interpreting the results for such dynamic properties is not possible without relying on the results of the preceding section on equilibrium properties.

\begin{figure}[t]
\centering
\includegraphics[width=1.0\linewidth]{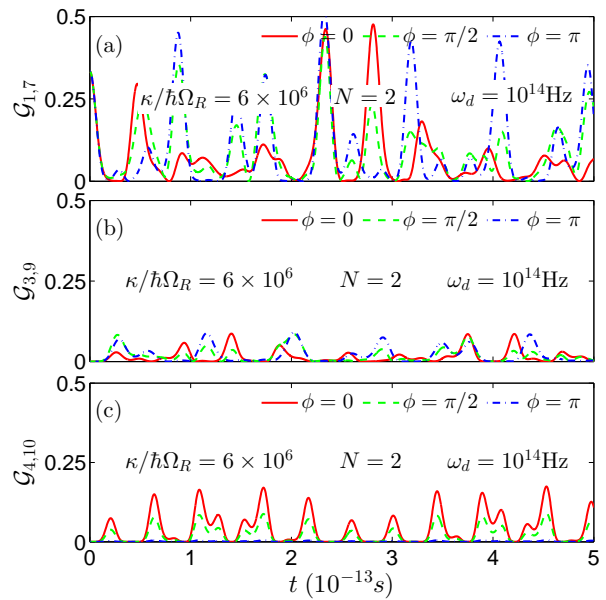}
\caption{(Color online) Two-point correlation functions of Eq.~(\ref{eq:corr_func}) associated with the propagation of an illustrative two-photon \emph{superposition} in Eq.~(\ref{eq:superposition}).}
\label{fig:superposition}
\end{figure}

The propagation of an illustrative delocalized \emph{superposition} of the form
\begin{eqnarray}
\label{eq:superposition}
\nonumber
|\Psi(t_0)\rangle & = & \frac{1}{\sqrt3}(|2\rangle |0\rangle^{\otimes 11}+|0\rangle^{\otimes 6}|2\rangle|0\rangle^{\otimes 5} \\
& & \qquad  +e^{i\phi}|1\rangle|0\rangle^{\otimes 5}|1\rangle|0\rangle^{\otimes 5}) \; ,
\end{eqnarray}
for various values of the relative phase $\phi$ is analyzed in Fig.~\ref{fig:superposition}. The strongest dialogue occurs between the cavities $(k,k')= (1,7)$ from which the photons emanate, as Fig.~\ref{fig:superposition}(a) shows. The relative phase $\phi$ can nonetheless affect the position of the peaks and the interference pattern can be engineered through tuning $\phi$. A strongly suppressed correlation profile is observed between cavities $(k,k')= (3,9)$ with an asymmetric alignment in the kagome as Fig.~\ref{fig:superposition}(b) illustrates. A fully regular and coherent correlation pattern finally emerges between cavities $(k,k')= (4,10)$ equally spaced from those the photons start out from. In this case, constructive and destructive interference patterns correspond to $\phi=0$ and $\pi$, respectively. The point has been illustrated in Fig.~\ref{fig:superposition}(c) and provides evidence for possible observation of \emph{resonant transmission} or Fabry-Pe\'rot-like resonances~\cite{Hecht1987} in such kagome interferometry device.

\section{Conclusions and outlook}
\label{sec:conclusions}

In conclusion, we have proposed a flexible numerical framework based on projected-entangled pair states (PEPS) for analyzing various static and dynamical properties of a photonic quantum simulator on a kagome geometry and have compared the results to the exact diagonalization (ED) method. A quantitative assessment of the role of undesired disorder in fabricating photonic quantum simulators has consequently become possible. The results confirm the robustness of the ground-state structure as well as two-point correlation functions associated with propagation of photonic excitations in the kagome. Our results promise, additionally, possible application of the cell as a tunable quantum interferometer with intriguing interference and resonant features.

We remind that the applicability and efficiency of the PEPS method in general is a nontrivial issue \emph{a priori} owing to the dependence of the entanglement entropy on the size of the system in two spatial dimensions unlike the one-dimensional analogue of the hierarchy of matrix-product states (MPS) (in which the latter stays constant~\cite{Eisert2010}). Our results justify the use of PEPS for addressing a prototypical photonic quantum simulator by providing numerical evidence in fair agreement with ED and promise thereby possible extension of the method for exploring arrays of larger sizes in the ultimate interest of further experimental developments toward scalable fabrication of photonic quantum simulators. Having said this, however, further progress in demonstrating the relative utility and power of PEPS beyond the applicability realm of ED will surely require access to more sophisticated computational resources than those exploited in the present work, and might deserve to be the subject of future investigations.

A variety of thrilling collective many-photon phenomena such as possible fermionization of photons and the emergence of a Tonks-Girardeau phase~\cite{DSouza2013}, formation of bound states of single photons, photon blockade, anomalous Hall effects~\cite{Petrescu2012} and other exotic phases of light may be explored upon a systematic extension of the proposed PEPS-based numerical machinery to larger kagome arrays. Simulation of the kagome lattice in the ultrastrong coupling regime of light-matter interaction~\cite{Niemczyk2010,Romero2012,Felicetti2013,Schiro2012} (the so-called Rabi-Hubbard paradigm of quantum optics~\cite{Schiro2012}) could be another exciting research avenue. One should note, however, that the Abelian $U(1)$ symmetry shall be broken down to the $\mathbb{Z}_2$ one in the presence of the counter-rotating terms in the ultrastrong coupling regime~\cite{Braak2011}. Finally, the application of the method to an ensemble of two-level atoms interacting with a bosonic mode described by the Dicke model is also another possibility to explore.

\begin{acknowledgments}

We acknowledge stimulating discussions with Guillermo Romero, Simone Felicetti, Mikel Sanz, Enrique Solano, Rom\'an Or\'us, Tom\'{a}\v{s} Opatrn\'y, Jalil Khatibi Moqaddam, and Reza Haghshenas.
We particularly would like to thank the referees for their constructive comments that led to significant improvements in the presentation of the materials.
H.S. is grateful to Universidad del Pa\'{\i}s Vasco and Aarhus University for support and hospitality. This work was financed by the European Social Fund and the state budget of the Czech Republic, project CZ.1.07/2.3.00/30.0041, and Sharif University of Technology's Office of Vice President for Research.

\end{acknowledgments}

\bibliography{Kagome_PEPS}

\begin{thebibliography}{34}%
\makeatletter
\providecommand \@ifxundefined [1]{%
 \@ifx{#1\undefined}
}%
\providecommand \@ifnum [1]{%
 \ifnum #1\expandafter \@firstoftwo
 \else \expandafter \@secondoftwo
 \fi
}%
\providecommand \@ifx [1]{%
 \ifx #1\expandafter \@firstoftwo
 \else \expandafter \@secondoftwo
 \fi
}%
\providecommand \natexlab [1]{#1}%
\providecommand \enquote  [1]{``#1''}%
\providecommand \bibnamefont  [1]{#1}%
\providecommand \bibfnamefont [1]{#1}%
\providecommand \citenamefont [1]{#1}%
\providecommand \href@noop [0]{\@secondoftwo}%
\providecommand \href [0]{\begingroup \@sanitize@url \@href}%
\providecommand \@href[1]{\@@startlink{#1}\@@href}%
\providecommand \@@href[1]{\endgroup#1\@@endlink}%
\providecommand \@sanitize@url [0]{\catcode `\\12\catcode `\$12\catcode
  `\&12\catcode `\#12\catcode `\^12\catcode `\_12\catcode `\%12\relax}%
\providecommand \@@startlink[1]{}%
\providecommand \@@endlink[0]{}%
\providecommand \url  [0]{\begingroup\@sanitize@url \@url }%
\providecommand \@url [1]{\endgroup\@href {#1}{\urlprefix }}%
\providecommand \urlprefix  [0]{URL }%
\providecommand \Eprint [0]{\href }%
\providecommand \doibase [0]{http://dx.doi.org/}%
\providecommand \selectlanguage [0]{\@gobble}%
\providecommand \bibinfo  [0]{\@secondoftwo}%
\providecommand \bibfield  [0]{\@secondoftwo}%
\providecommand \translation [1]{[#1]}%
\providecommand \BibitemOpen [0]{}%
\providecommand \bibitemStop [0]{}%
\providecommand \bibitemNoStop [0]{.\EOS\space}%
\providecommand \EOS [0]{\spacefactor3000\relax}%
\providecommand \BibitemShut  [1]{\csname bibitem#1\endcsname}%
\let\auto@bib@innerbib\@empty
\bibitem [{\citenamefont {Feynman}(1982)}]{Feynman1982}%
  \BibitemOpen
  \bibfield  {author} {\bibinfo {author} {\bibfnamefont {R.~P.}\ \bibnamefont
  {Feynman}},\ }\href {\doibase 10.1007/BF02650179} {\bibfield  {journal}
  {\bibinfo  {journal} {Int. J. Theor. Phys.}\ }\textbf {\bibinfo {volume}
  {21}},\ \bibinfo {pages} {467} (\bibinfo {year} {1982})}\BibitemShut
  {NoStop}%
\bibitem [{\citenamefont {Houck}\ \emph {et~al.}(2012)\citenamefont {Houck},
  \citenamefont {T{\"u}reci},\ and\ \citenamefont {Koch}}]{Houck2012}%
  \BibitemOpen
  \bibfield  {author} {\bibinfo {author} {\bibfnamefont {A.}~\bibnamefont
  {Houck}}, \bibinfo {author} {\bibfnamefont {H.}~\bibnamefont {T{\"u}reci}}, \
  and\ \bibinfo {author} {\bibfnamefont {J.}~\bibnamefont {Koch}},\ }\href
  {\doibase 10.1038/nphys2251} {\bibfield  {journal} {\bibinfo  {journal}
  {Nature Phys.}\ }\textbf {\bibinfo {volume} {8}},\ \bibinfo {pages} {292}
  (\bibinfo {year} {2012})}\BibitemShut {NoStop}%
\bibitem [{\citenamefont {Hartmann}\ \emph {et~al.}(2006)\citenamefont
  {Hartmann}, \citenamefont {Brand\~{a}o},\ and\ \citenamefont
  {Plenio}}]{Hartmann2006}%
  \BibitemOpen
  \bibfield  {author} {\bibinfo {author} {\bibfnamefont {M.~J.}\ \bibnamefont
  {Hartmann}}, \bibinfo {author} {\bibfnamefont {F.~G.}\ \bibnamefont
  {Brand\~{a}o}}, \ and\ \bibinfo {author} {\bibfnamefont {M.~B.}\ \bibnamefont
  {Plenio}},\ }\href {\doibase 10.1038/nphys462} {\bibfield  {journal}
  {\bibinfo  {journal} {Nature Phys.}\ }\textbf {\bibinfo {volume} {2}},\
  \bibinfo {pages} {849} (\bibinfo {year} {2006})}\BibitemShut {NoStop}%
\bibitem [{\citenamefont {Greentree}\ \emph {et~al.}(2006)\citenamefont
  {Greentree}, \citenamefont {Tahan}, \citenamefont {Cole},\ and\ \citenamefont
  {Hollenberg}}]{Greentree2006}%
  \BibitemOpen
  \bibfield  {author} {\bibinfo {author} {\bibfnamefont {A.~D.}\ \bibnamefont
  {Greentree}}, \bibinfo {author} {\bibfnamefont {C.}~\bibnamefont {Tahan}},
  \bibinfo {author} {\bibfnamefont {J.~H.}\ \bibnamefont {Cole}}, \ and\
  \bibinfo {author} {\bibfnamefont {L.~C.}\ \bibnamefont {Hollenberg}},\ }\href
  {\doibase 10.1038/nphys466} {\bibfield  {journal} {\bibinfo  {journal}
  {Nature Phys.}\ }\textbf {\bibinfo {volume} {2}},\ \bibinfo {pages} {856}
  (\bibinfo {year} {2006})}\BibitemShut {NoStop}%
\bibitem [{\citenamefont {Koch}\ and\ \citenamefont {Le~Hur}(2009)}]{Koch2009}%
  \BibitemOpen
  \bibfield  {author} {\bibinfo {author} {\bibfnamefont {J.}~\bibnamefont
  {Koch}}\ and\ \bibinfo {author} {\bibfnamefont {K.}~\bibnamefont {Le~Hur}},\
  }\href {\doibase 10.1103/PhysRevA.80.023811} {\bibfield  {journal} {\bibinfo
  {journal} {Phys. Rev. A}\ }\textbf {\bibinfo {volume} {80}},\ \bibinfo
  {pages} {023811} (\bibinfo {year} {2009})}\BibitemShut {NoStop}%
\bibitem [{\citenamefont {D'Souza}\ \emph {et~al.}(2013)\citenamefont
  {D'Souza}, \citenamefont {Sanders},\ and\ \citenamefont
  {Feder}}]{DSouza2013}%
  \BibitemOpen
  \bibfield  {author} {\bibinfo {author} {\bibfnamefont {A.~G.}\ \bibnamefont
  {D'Souza}}, \bibinfo {author} {\bibfnamefont {B.~C.}\ \bibnamefont
  {Sanders}}, \ and\ \bibinfo {author} {\bibfnamefont {D.~L.}\ \bibnamefont
  {Feder}},\ }\href {\doibase 10.1103/PhysRevA.88.063801} {\bibfield  {journal}
  {\bibinfo  {journal} {Phys. Rev. A}\ }\textbf {\bibinfo {volume} {88}},\
  \bibinfo {pages} {063801} (\bibinfo {year} {2013})}\BibitemShut {NoStop}%
\bibitem [{\citenamefont {Petrescu}\ \emph {et~al.}(2012)\citenamefont
  {Petrescu}, \citenamefont {Houck},\ and\ \citenamefont
  {Le~Hur}}]{Petrescu2012}%
  \BibitemOpen
  \bibfield  {author} {\bibinfo {author} {\bibfnamefont {A.}~\bibnamefont
  {Petrescu}}, \bibinfo {author} {\bibfnamefont {A.~A.}\ \bibnamefont {Houck}},
  \ and\ \bibinfo {author} {\bibfnamefont {K.}~\bibnamefont {Le~Hur}},\ }\href
  {\doibase 10.1103/PhysRevA.86.053804} {\bibfield  {journal} {\bibinfo
  {journal} {Phys. Rev. A}\ }\textbf {\bibinfo {volume} {86}},\ \bibinfo
  {pages} {053804} (\bibinfo {year} {2012})}\BibitemShut {NoStop}%
\bibitem [{\citenamefont {Schmidt}\ \emph {et~al.}(2010)\citenamefont
  {Schmidt}, \citenamefont {Gerace}, \citenamefont {Houck}, \citenamefont
  {Blatter},\ and\ \citenamefont {T\"ureci}}]{Schmidt2010}%
  \BibitemOpen
  \bibfield  {author} {\bibinfo {author} {\bibfnamefont {S.}~\bibnamefont
  {Schmidt}}, \bibinfo {author} {\bibfnamefont {D.}~\bibnamefont {Gerace}},
  \bibinfo {author} {\bibfnamefont {A.~A.}\ \bibnamefont {Houck}}, \bibinfo
  {author} {\bibfnamefont {G.}~\bibnamefont {Blatter}}, \ and\ \bibinfo
  {author} {\bibfnamefont {H.~E.}\ \bibnamefont {T\"ureci}},\ }\href {\doibase
  10.1103/PhysRevB.82.100507} {\bibfield  {journal} {\bibinfo  {journal} {Phys.
  Rev. B}\ }\textbf {\bibinfo {volume} {82}},\ \bibinfo {pages} {100507}
  (\bibinfo {year} {2010})}\BibitemShut {NoStop}%
\bibitem [{\citenamefont {Underwood}\ \emph {et~al.}(2012)\citenamefont
  {Underwood}, \citenamefont {Shanks}, \citenamefont {Koch},\ and\
  \citenamefont {Houck}}]{Underwood2012}%
  \BibitemOpen
  \bibfield  {author} {\bibinfo {author} {\bibfnamefont {D.~L.}\ \bibnamefont
  {Underwood}}, \bibinfo {author} {\bibfnamefont {W.~E.}\ \bibnamefont
  {Shanks}}, \bibinfo {author} {\bibfnamefont {J.}~\bibnamefont {Koch}}, \ and\
  \bibinfo {author} {\bibfnamefont {A.~A.}\ \bibnamefont {Houck}},\ }\href
  {\doibase 10.1103/PhysRevA.86.023837} {\bibfield  {journal} {\bibinfo
  {journal} {Phys. Rev. A}\ }\textbf {\bibinfo {volume} {86}},\ \bibinfo
  {pages} {023837} (\bibinfo {year} {2012})}\BibitemShut {NoStop}%
\bibitem [{\citenamefont {Yan}\ \emph {et~al.}(2011)\citenamefont {Yan},
  \citenamefont {Huse},\ and\ \citenamefont {White}}]{Yan2011}%
  \BibitemOpen
  \bibfield  {author} {\bibinfo {author} {\bibfnamefont {S.}~\bibnamefont
  {Yan}}, \bibinfo {author} {\bibfnamefont {D.~A.}\ \bibnamefont {Huse}}, \
  and\ \bibinfo {author} {\bibfnamefont {S.~R.}\ \bibnamefont {White}},\ }\href
  {\doibase 10.1126/science.1201080} {\bibfield  {journal} {\bibinfo  {journal}
  {Science}\ }\textbf {\bibinfo {volume} {332}},\ \bibinfo {pages} {1173}
  (\bibinfo {year} {2011})}\BibitemShut {NoStop}%
\bibitem [{\citenamefont {Angelakis}\ \emph {et~al.}(2007)\citenamefont
  {Angelakis}, \citenamefont {Santos},\ and\ \citenamefont
  {Bose}}]{Angelakis2007}%
  \BibitemOpen
  \bibfield  {author} {\bibinfo {author} {\bibfnamefont {D.~G.}\ \bibnamefont
  {Angelakis}}, \bibinfo {author} {\bibfnamefont {M.~F.}\ \bibnamefont
  {Santos}}, \ and\ \bibinfo {author} {\bibfnamefont {S.}~\bibnamefont
  {Bose}},\ }\href {\doibase 10.1103/PhysRevA.76.031805} {\bibfield  {journal}
  {\bibinfo  {journal} {Phys. Rev. A}\ }\textbf {\bibinfo {volume} {76}},\
  \bibinfo {pages} {031805} (\bibinfo {year} {2007})}\BibitemShut {NoStop}%
\bibitem [{\citenamefont {Hartmann}(2010)}]{Hartmann2012}%
  \BibitemOpen
  \bibfield  {author} {\bibinfo {author} {\bibfnamefont {M.~J.}\ \bibnamefont
  {Hartmann}},\ }\href {\doibase 10.1103/PhysRevLett.104.113601} {\bibfield
  {journal} {\bibinfo  {journal} {Phys. Rev. Lett.}\ }\textbf {\bibinfo
  {volume} {104}},\ \bibinfo {pages} {113601} (\bibinfo {year}
  {2010})}\BibitemShut {NoStop}%
\bibitem [{\citenamefont {Schir\'o}\ \emph {et~al.}(2012)\citenamefont
  {Schir\'o}, \citenamefont {Bordyuh}, \citenamefont {\"Oztop},\ and\
  \citenamefont {T\"ureci}}]{Schiro2012}%
  \BibitemOpen
  \bibfield  {author} {\bibinfo {author} {\bibfnamefont {M.}~\bibnamefont
  {Schir\'o}}, \bibinfo {author} {\bibfnamefont {M.}~\bibnamefont {Bordyuh}},
  \bibinfo {author} {\bibfnamefont {B.}~\bibnamefont {\"Oztop}}, \ and\
  \bibinfo {author} {\bibfnamefont {H.~E.}\ \bibnamefont {T\"ureci}},\ }\href
  {\doibase 10.1103/PhysRevLett.109.053601} {\bibfield  {journal} {\bibinfo
  {journal} {Phys. Rev. Lett.}\ }\textbf {\bibinfo {volume} {109}},\ \bibinfo
  {pages} {053601} (\bibinfo {year} {2012})}\BibitemShut {NoStop}%
\bibitem [{\citenamefont {Zheng}\ and\ \citenamefont
  {Takada}(2011)}]{Zheng2011}%
  \BibitemOpen
  \bibfield  {author} {\bibinfo {author} {\bibfnamefont {H.}~\bibnamefont
  {Zheng}}\ and\ \bibinfo {author} {\bibfnamefont {Y.}~\bibnamefont {Takada}},\
  }\href {\doibase 10.1103/PhysRevA.84.043819} {\bibfield  {journal} {\bibinfo
  {journal} {Phys. Rev. A}\ }\textbf {\bibinfo {volume} {84}},\ \bibinfo
  {pages} {043819} (\bibinfo {year} {2011})}\BibitemShut {NoStop}%
\bibitem [{\citenamefont {Sch\"on}\ \emph {et~al.}(2005)\citenamefont
  {Sch\"on}, \citenamefont {Solano}, \citenamefont {Verstraete}, \citenamefont
  {Cirac},\ and\ \citenamefont {Wolf}}]{Schoen2005}%
  \BibitemOpen
  \bibfield  {author} {\bibinfo {author} {\bibfnamefont {C.}~\bibnamefont
  {Sch\"on}}, \bibinfo {author} {\bibfnamefont {E.}~\bibnamefont {Solano}},
  \bibinfo {author} {\bibfnamefont {F.}~\bibnamefont {Verstraete}}, \bibinfo
  {author} {\bibfnamefont {J.~I.}\ \bibnamefont {Cirac}}, \ and\ \bibinfo
  {author} {\bibfnamefont {M.~M.}\ \bibnamefont {Wolf}},\ }\href {\doibase
  10.1103/PhysRevLett.95.110503} {\bibfield  {journal} {\bibinfo  {journal}
  {Phys. Rev. Lett.}\ }\textbf {\bibinfo {volume} {95}},\ \bibinfo {eid}
  {110503} (\bibinfo {year} {2005})}\BibitemShut {NoStop}%
\bibitem [{\citenamefont {Perez-Garcia}\ \emph {et~al.}(2007)\citenamefont
  {Perez-Garcia}, \citenamefont {Verstraete}, \citenamefont {Wolf},\ and\
  \citenamefont {Cirac}}]{Perez2007}%
  \BibitemOpen
  \bibfield  {author} {\bibinfo {author} {\bibfnamefont {D.}~\bibnamefont
  {Perez-Garcia}}, \bibinfo {author} {\bibfnamefont {F.}~\bibnamefont
  {Verstraete}}, \bibinfo {author} {\bibfnamefont {M.~M.}\ \bibnamefont
  {Wolf}}, \ and\ \bibinfo {author} {\bibfnamefont {J.~I.}\ \bibnamefont
  {Cirac}},\ }\href@noop {} {\bibfield  {journal} {\bibinfo  {journal} {Quantum
  Inf. Comput.}\ }\textbf {\bibinfo {volume} {7}},\ \bibinfo {pages} {401}
  (\bibinfo {year} {2007})}\BibitemShut {NoStop}%
\bibitem [{\citenamefont {Saberi}\ \emph {et~al.}(2009)\citenamefont {Saberi},
  \citenamefont {Weichselbaum}, \citenamefont {Lamata}, \citenamefont
  {Perez-Garcia}, \citenamefont {von Delft},\ and\ \citenamefont
  {Solano}}]{Saberi2009}%
  \BibitemOpen
  \bibfield  {author} {\bibinfo {author} {\bibfnamefont {H.}~\bibnamefont
  {Saberi}}, \bibinfo {author} {\bibfnamefont {A.}~\bibnamefont
  {Weichselbaum}}, \bibinfo {author} {\bibfnamefont {L.}~\bibnamefont
  {Lamata}}, \bibinfo {author} {\bibfnamefont {D.}~\bibnamefont
  {Perez-Garcia}}, \bibinfo {author} {\bibfnamefont {J.}~\bibnamefont {von
  Delft}}, \ and\ \bibinfo {author} {\bibfnamefont {E.}~\bibnamefont
  {Solano}},\ }\href {\doibase 10.1103/PhysRevA.80.022334} {\bibfield
  {journal} {\bibinfo  {journal} {Phys. Rev. A}\ }\textbf {\bibinfo {volume}
  {80}},\ \bibinfo {pages} {022334} (\bibinfo {year} {2009})}\BibitemShut
  {NoStop}%
\bibitem [{\citenamefont {Saberi}\ and\ \citenamefont
  {Mardoukhi}(2012)}]{Saberi2012}%
  \BibitemOpen
  \bibfield  {author} {\bibinfo {author} {\bibfnamefont {H.}~\bibnamefont
  {Saberi}}\ and\ \bibinfo {author} {\bibfnamefont {Y.}~\bibnamefont
  {Mardoukhi}},\ }\href {\doibase 10.1103/PhysRevA.85.052323} {\bibfield
  {journal} {\bibinfo  {journal} {Phys. Rev. A}\ }\textbf {\bibinfo {volume}
  {85}},\ \bibinfo {pages} {052323} (\bibinfo {year} {2012})}\BibitemShut
  {NoStop}%
\bibitem [{\citenamefont {Verstraete}\ \emph {et~al.}(2008)\citenamefont
  {Verstraete}, \citenamefont {Murg},\ and\ \citenamefont
  {Cirac}}]{Verstraete2008}%
  \BibitemOpen
  \bibfield  {author} {\bibinfo {author} {\bibfnamefont {F.}~\bibnamefont
  {Verstraete}}, \bibinfo {author} {\bibfnamefont {V.}~\bibnamefont {Murg}}, \
  and\ \bibinfo {author} {\bibfnamefont {J.}~\bibnamefont {Cirac}},\ }\href
  {\doibase 10.1080/14789940801912366} {\bibfield  {journal} {\bibinfo
  {journal} {Adv. Phys.}\ }\textbf {\bibinfo {volume} {57}},\ \bibinfo {pages}
  {143} (\bibinfo {year} {2008})}\BibitemShut {NoStop}%
\bibitem [{\citenamefont {Schuch}\ \emph {et~al.}(2010)\citenamefont {Schuch},
  \citenamefont {Cirac},\ and\ \citenamefont
  {P{\'e}rez-Garc{\'\i}a}}]{Schuch2010}%
  \BibitemOpen
  \bibfield  {author} {\bibinfo {author} {\bibfnamefont {N.}~\bibnamefont
  {Schuch}}, \bibinfo {author} {\bibfnamefont {I.}~\bibnamefont {Cirac}}, \
  and\ \bibinfo {author} {\bibfnamefont {D.}~\bibnamefont
  {P{\'e}rez-Garc{\'\i}a}},\ }\href {\doibase 10.1016/j.aop.2010.05.008}
  {\bibfield  {journal} {\bibinfo  {journal} {Ann. Phys.}\ }\textbf {\bibinfo
  {volume} {325}},\ \bibinfo {pages} {2153} (\bibinfo {year}
  {2010})}\BibitemShut {NoStop}%
\bibitem [{\citenamefont {Hoffman}\ \emph {et~al.}(2011)\citenamefont
  {Hoffman}, \citenamefont {Srinivasan}, \citenamefont {Schmidt}, \citenamefont
  {Spietz}, \citenamefont {Aumentado}, \citenamefont {T\"ureci},\ and\
  \citenamefont {Houck}}]{Hoffman2011}%
  \BibitemOpen
  \bibfield  {author} {\bibinfo {author} {\bibfnamefont {A.~J.}\ \bibnamefont
  {Hoffman}}, \bibinfo {author} {\bibfnamefont {S.~J.}\ \bibnamefont
  {Srinivasan}}, \bibinfo {author} {\bibfnamefont {S.}~\bibnamefont {Schmidt}},
  \bibinfo {author} {\bibfnamefont {L.}~\bibnamefont {Spietz}}, \bibinfo
  {author} {\bibfnamefont {J.}~\bibnamefont {Aumentado}}, \bibinfo {author}
  {\bibfnamefont {H.~E.}\ \bibnamefont {T\"ureci}}, \ and\ \bibinfo {author}
  {\bibfnamefont {A.~A.}\ \bibnamefont {Houck}},\ }\href {\doibase
  10.1103/PhysRevLett.107.053602} {\bibfield  {journal} {\bibinfo  {journal}
  {Phys. Rev. Lett.}\ }\textbf {\bibinfo {volume} {107}},\ \bibinfo {pages}
  {053602} (\bibinfo {year} {2011})}\BibitemShut {NoStop}%
\bibitem [{\citenamefont {Orus}(2013)}]{Orus2013}%
  \BibitemOpen
  \bibfield  {author} {\bibinfo {author} {\bibfnamefont {R.}~\bibnamefont
  {Orus}},\ }\href@noop {} {\bibfield  {journal} {\bibinfo  {journal}
  {arXiv:1306.2164}\ } (\bibinfo {year} {2013})}\BibitemShut {NoStop}%
\bibitem [{\citenamefont {Verstraete}\ \emph {et~al.}(2004)\citenamefont
  {Verstraete}, \citenamefont {Porras},\ and\ \citenamefont
  {Cirac}}]{Verstraete2004}%
  \BibitemOpen
  \bibfield  {author} {\bibinfo {author} {\bibfnamefont {F.}~\bibnamefont
  {Verstraete}}, \bibinfo {author} {\bibfnamefont {D.}~\bibnamefont {Porras}},
  \ and\ \bibinfo {author} {\bibfnamefont {J.~I.}\ \bibnamefont {Cirac}},\
  }\href {\doibase 10.1103/PhysRevLett.93.227205} {\bibfield  {journal}
  {\bibinfo  {journal} {Phys. Rev. Lett.}\ }\textbf {\bibinfo {volume} {93}},\
  \bibinfo {pages} {227205} (\bibinfo {year} {2004})}\BibitemShut {NoStop}%
\bibitem [{\citenamefont {Saberi}\ \emph {et~al.}(2008)\citenamefont {Saberi},
  \citenamefont {Weichselbaum},\ and\ \citenamefont {von Delft}}]{Saberi2008}%
  \BibitemOpen
  \bibfield  {author} {\bibinfo {author} {\bibfnamefont {H.}~\bibnamefont
  {Saberi}}, \bibinfo {author} {\bibfnamefont {A.}~\bibnamefont
  {Weichselbaum}}, \ and\ \bibinfo {author} {\bibfnamefont {J.}~\bibnamefont
  {von Delft}},\ }\href {\doibase 10.1103/PhysRevB.78.035124} {\bibfield
  {journal} {\bibinfo  {journal} {Phys. Rev. B}\ }\textbf {\bibinfo {volume}
  {78}},\ \bibinfo {eid} {035124} (\bibinfo {year} {2008})}\BibitemShut
  {NoStop}%
\bibitem [{\citenamefont {Bardyn}\ and\ \citenamefont {\ifmmode \dot{I}\else
  \.{I}\fi{}mamo\ifmmode~\check{g}\else \v{g}\fi{}lu}(2012)}]{Bardyn2012}%
  \BibitemOpen
  \bibfield  {author} {\bibinfo {author} {\bibfnamefont {C.-E.}\ \bibnamefont
  {Bardyn}}\ and\ \bibinfo {author} {\bibfnamefont {A.}~\bibnamefont {\ifmmode
  \dot{I}\else \.{I}\fi{}mamo\ifmmode~\check{g}\else \v{g}\fi{}lu}},\ }\href
  {\doibase 10.1103/PhysRevLett.109.253606} {\bibfield  {journal} {\bibinfo
  {journal} {Phys. Rev. Lett.}\ }\textbf {\bibinfo {volume} {109}},\ \bibinfo
  {pages} {253606} (\bibinfo {year} {2012})}\BibitemShut {NoStop}%
\bibitem [{\citenamefont {Bauer}\ \emph {et~al.}(2011)\citenamefont {Bauer},
  \citenamefont {Corboz}, \citenamefont {Or\'us},\ and\ \citenamefont
  {Troyer}}]{Bauer2011}%
  \BibitemOpen
  \bibfield  {author} {\bibinfo {author} {\bibfnamefont {B.}~\bibnamefont
  {Bauer}}, \bibinfo {author} {\bibfnamefont {P.}~\bibnamefont {Corboz}},
  \bibinfo {author} {\bibfnamefont {R.}~\bibnamefont {Or\'us}}, \ and\ \bibinfo
  {author} {\bibfnamefont {M.}~\bibnamefont {Troyer}},\ }\href {\doibase
  10.1103/PhysRevB.83.125106} {\bibfield  {journal} {\bibinfo  {journal} {Phys.
  Rev. B}\ }\textbf {\bibinfo {volume} {83}},\ \bibinfo {pages} {125106}
  (\bibinfo {year} {2011})}\BibitemShut {NoStop}%
\bibitem [{\citenamefont {Wallraff}\ \emph {et~al.}(2004)\citenamefont
  {Wallraff}, \citenamefont {Schuster}, \citenamefont {Blais}, \citenamefont
  {Frunzio}, \citenamefont {Huang}, \citenamefont {Majer}, \citenamefont
  {Kumar}, \citenamefont {Girvin},\ and\ \citenamefont
  {Schoelkopf}}]{Wallraff2004}%
  \BibitemOpen
  \bibfield  {author} {\bibinfo {author} {\bibfnamefont {A.}~\bibnamefont
  {Wallraff}}, \bibinfo {author} {\bibfnamefont {D.}~\bibnamefont {Schuster}},
  \bibinfo {author} {\bibfnamefont {A.}~\bibnamefont {Blais}}, \bibinfo
  {author} {\bibfnamefont {L.}~\bibnamefont {Frunzio}}, \bibinfo {author}
  {\bibfnamefont {R.}~\bibnamefont {Huang}}, \bibinfo {author} {\bibfnamefont
  {J.}~\bibnamefont {Majer}}, \bibinfo {author} {\bibfnamefont
  {S.}~\bibnamefont {Kumar}}, \bibinfo {author} {\bibfnamefont
  {S.}~\bibnamefont {Girvin}}, \ and\ \bibinfo {author} {\bibfnamefont
  {R.}~\bibnamefont {Schoelkopf}},\ }\href {\doibase 10.1038/nature02851}
  {\bibfield  {journal} {\bibinfo  {journal} {Nature}\ }\textbf {\bibinfo
  {volume} {431}},\ \bibinfo {pages} {162} (\bibinfo {year}
  {2004})}\BibitemShut {NoStop}%
\bibitem [{\citenamefont {Niemczyk}\ \emph {et~al.}(2010)\citenamefont
  {Niemczyk}, \citenamefont {Deppe}, \citenamefont {Huebl}, \citenamefont
  {Menzel}, \citenamefont {Hocke}, \citenamefont {Schwarz}, \citenamefont
  {Garcia-Ripoll}, \citenamefont {Zueco}, \citenamefont {H{\"u}mmer},
  \citenamefont {Solano}, \citenamefont {Marx},\ and\ \citenamefont
  {R.}}]{Niemczyk2010}%
  \BibitemOpen
  \bibfield  {author} {\bibinfo {author} {\bibfnamefont {T.}~\bibnamefont
  {Niemczyk}}, \bibinfo {author} {\bibfnamefont {F.}~\bibnamefont {Deppe}},
  \bibinfo {author} {\bibfnamefont {H.}~\bibnamefont {Huebl}}, \bibinfo
  {author} {\bibfnamefont {E.}~\bibnamefont {Menzel}}, \bibinfo {author}
  {\bibfnamefont {F.}~\bibnamefont {Hocke}}, \bibinfo {author} {\bibfnamefont
  {M.}~\bibnamefont {Schwarz}}, \bibinfo {author} {\bibfnamefont
  {J.}~\bibnamefont {Garcia-Ripoll}}, \bibinfo {author} {\bibfnamefont
  {D.}~\bibnamefont {Zueco}}, \bibinfo {author} {\bibfnamefont
  {T.}~\bibnamefont {H{\"u}mmer}}, \bibinfo {author} {\bibfnamefont
  {E.}~\bibnamefont {Solano}}, \bibinfo {author} {\bibfnamefont
  {A.}~\bibnamefont {Marx}}, \ and\ \bibinfo {author} {\bibfnamefont
  {G.}~\bibnamefont {R.}},\ }\href {\doibase 10.1038/nphys1730} {\bibfield
  {journal} {\bibinfo  {journal} {Nature Phys.}\ }\textbf {\bibinfo {volume}
  {6}},\ \bibinfo {pages} {772} (\bibinfo {year} {2010})}\BibitemShut {NoStop}%
\bibitem [{\citenamefont {Romero}\ \emph {et~al.}(2012)\citenamefont {Romero},
  \citenamefont {Ballester}, \citenamefont {Wang}, \citenamefont {Scarani},\
  and\ \citenamefont {Solano}}]{Romero2012}%
  \BibitemOpen
  \bibfield  {author} {\bibinfo {author} {\bibfnamefont {G.}~\bibnamefont
  {Romero}}, \bibinfo {author} {\bibfnamefont {D.}~\bibnamefont {Ballester}},
  \bibinfo {author} {\bibfnamefont {Y.~M.}\ \bibnamefont {Wang}}, \bibinfo
  {author} {\bibfnamefont {V.}~\bibnamefont {Scarani}}, \ and\ \bibinfo
  {author} {\bibfnamefont {E.}~\bibnamefont {Solano}},\ }\href {\doibase
  10.1103/PhysRevLett.108.120501} {\bibfield  {journal} {\bibinfo  {journal}
  {Phys. Rev. Lett.}\ }\textbf {\bibinfo {volume} {108}},\ \bibinfo {pages}
  {120501} (\bibinfo {year} {2012})}\BibitemShut {NoStop}%
\bibitem [{\citenamefont {Felicetti}\ \emph {et~al.}(2013)\citenamefont
  {Felicetti}, \citenamefont {Romero}, \citenamefont {Rossini}, \citenamefont
  {Fazio},\ and\ \citenamefont {Solano}}]{Felicetti2013}%
  \BibitemOpen
  \bibfield  {author} {\bibinfo {author} {\bibfnamefont {S.}~\bibnamefont
  {Felicetti}}, \bibinfo {author} {\bibfnamefont {G.}~\bibnamefont {Romero}},
  \bibinfo {author} {\bibfnamefont {D.}~\bibnamefont {Rossini}}, \bibinfo
  {author} {\bibfnamefont {R.}~\bibnamefont {Fazio}}, \ and\ \bibinfo {author}
  {\bibfnamefont {E.}~\bibnamefont {Solano}},\ }\href@noop {} {\bibfield
  {journal} {\bibinfo  {journal} {arXiv:1304.6221}\ } (\bibinfo {year}
  {2013})}\BibitemShut {NoStop}%
\bibitem [{\citenamefont {De~Chiara}\ \emph {et~al.}(2012)\citenamefont
  {De~Chiara}, \citenamefont {Lepori}, \citenamefont {Lewenstein},\ and\
  \citenamefont {Sanpera}}]{DeChiara2012}%
  \BibitemOpen
  \bibfield  {author} {\bibinfo {author} {\bibfnamefont {G.}~\bibnamefont
  {De~Chiara}}, \bibinfo {author} {\bibfnamefont {L.}~\bibnamefont {Lepori}},
  \bibinfo {author} {\bibfnamefont {M.}~\bibnamefont {Lewenstein}}, \ and\
  \bibinfo {author} {\bibfnamefont {A.}~\bibnamefont {Sanpera}},\ }\href
  {\doibase 10.1103/PhysRevLett.109.237208} {\bibfield  {journal} {\bibinfo
  {journal} {Phys. Rev. Lett.}\ }\textbf {\bibinfo {volume} {109}},\ \bibinfo
  {pages} {237208} (\bibinfo {year} {2012})}\BibitemShut {NoStop}%
\bibitem [{\citenamefont {Hecht}(1987)}]{Hecht1987}%
  \BibitemOpen
  \bibfield  {author} {\bibinfo {author} {\bibfnamefont {E.}~\bibnamefont
  {Hecht}},\ }\href@noop {} {\emph {\bibinfo {title} {Optics}}}\ (\bibinfo
  {publisher} {Addison-Wesley, Reading},\ \bibinfo {year} {1987})\BibitemShut
  {NoStop}%
\bibitem [{\citenamefont {Eisert}\ \emph {et~al.}(2010)\citenamefont {Eisert},
  \citenamefont {Cramer},\ and\ \citenamefont {Plenio}}]{Eisert2010}%
  \BibitemOpen
  \bibfield  {author} {\bibinfo {author} {\bibfnamefont {J.}~\bibnamefont
  {Eisert}}, \bibinfo {author} {\bibfnamefont {M.}~\bibnamefont {Cramer}}, \
  and\ \bibinfo {author} {\bibfnamefont {M.~B.}\ \bibnamefont {Plenio}},\
  }\href {\doibase 10.1103/RevModPhys.82.277} {\bibfield  {journal} {\bibinfo
  {journal} {Rev. Mod. Phys.}\ }\textbf {\bibinfo {volume} {82}},\ \bibinfo
  {pages} {277} (\bibinfo {year} {2010})}\BibitemShut {NoStop}%
\bibitem [{\citenamefont {Braak}(2011)}]{Braak2011}%
  \BibitemOpen
  \bibfield  {author} {\bibinfo {author} {\bibfnamefont {D.}~\bibnamefont
  {Braak}},\ }\href {\doibase 10.1103/PhysRevLett.107.100401} {\bibfield
  {journal} {\bibinfo  {journal} {Phys. Rev. Lett.}\ }\textbf {\bibinfo
  {volume} {107}},\ \bibinfo {pages} {100401} (\bibinfo {year}
  {2011})}\BibitemShut {NoStop}%
\end{thebibliography}%

\end{document}